\documentclass[aps,pre,twocolumn]{revtex4}
\usepackage{amssymb}
\usepackage{amsmath}
\usepackage{bm}
\usepackage{mathtools}

\newcommand{\bigdave}[1]{\big\langle\hspace{-1.0mm}\big\langle#1
                          \big\rangle\hspace{-1.0mm}\big\rangle}
\newcommand{\Bigdave}[1]{\Big\langle\hspace{-1.4mm}\Big\langle#1
                          \Big\rangle\hspace{-1.4mm}\Big\rangle}

\DeclareMathOperator{\sign}{sign}

\begin{document}

\title{Persymmetric Jacobi matrices with square-integer eigenvalues\\
       and dispersionless mass-spring chains}

\author{Ruggero Vaia}
\affiliation{Istituto dei Sistemi Complessi, Consiglio Nazionale delle Ricerche,
         via Madonna del Piano 10, I-50019 Sesto Fiorentino, Italy}
\affiliation{Istituto Nazionale di Fisica Nucleare, Sezione di Firenze,
         via Giovanni Sansone 1, I-50019 Sesto Fiorentino, Italy}

\author{Lidia Spadini}
\affiliation{Dipartimento di Matematica e Informatica ``Ulisse Dini'',
             Universit\`a di Firenze,
             viale Morgagni 67/a, I-50134 Firenze, Italy}

\begin{abstract}
A real persymmetric Jacobi matrix of order $n$ whose eigenvalues are $\big\{2k^2\big\}_{k=0}^{n-1}$ is presented, with entries given as explicit functions of $n$. Besides the possible use for testing forward and inverse numerical algorithms, such a matrix is especially relevant for its connection with the dynamics of a mass-spring chain, which is a multi-purpose prototype model. Indeed, the mode frequencies being the square roots of the eigenvalues of the interaction matrix, one can shape the chain in such a way that its dynamics be perfectly periodic and dispersionless.
\end{abstract}

\maketitle

\section{Introduction}
\label{s.intro}

The task of finding the matrix elements of a tridiagonal bisymmetric (i.e., symmetric and persymmetric) matrix of order $n$ starting from the requirement that it have a given spectrum $\{\lambda_k\}_{k=0}^{n-1}$ is a well-posed `inverse problem'.
In particular, the number of independent matrix elements to be determined is equal to its dimension $n$ and hence to the number of given eigenvalues.
A very efficient algorithm, aimed at a numerical approach, was developed by de\,Boor and Golub~\cite{deBoorG1978}: it is based on a sequence of $n{+}1$ monic polynomials $\{\chi_i(\lambda)\}_{i=0}^n$ of degree $i$, corresponding to the characteristic polynomials of the matrix $\bm{A}$ and of its submatrices, which are orthogonal under a suitable internal product defined in terms of the given eigenvalues; thanks to orthogonality the polynomials can be constructed sequentially, starting from $\chi_0(\lambda)=1$, at the same time yielding the matrix elements.

Even though the usefulness of the de\,Boor-Golub algorithm is almost confined to the numerical side~\cite{BrudererFRBO2012}, it may happen that the internal product can be dealt with analytically. This is the case, for instance, when the eigenvalues are the sequence of the first $n$ integers, $\lambda_k\propto{k}$. Here it is shown that the case $\lambda_k\propto{k^2}$ can be also managed, yielding to analytic expressions for the first few matrix elements, as shown in \ref{a.analytic}. Even though this analytic approach cannot be pursued to obtain all matrix entries, it provided hints leading to guess analytic expressions for the matrix elements, valid for any dimension $n$, whose proof is the main result provided here.

Knowing a matrix with eigenvalues $\lambda_k\propto{k^2}$ is relevant because of its connection with the mass-spring chain model~\cite{Vaia2018}, constituted by a sequence of $n$ massive bodies pairwise connected by ideal springs with given elastic constants. Such a prototype model has a wide generality in Physics and Engineering. Thanks to the results presented here, it is possible to design the sequence of masses and elastic constants in such a way that the chain dynamics be exactly periodic. Indeed, the square roots of the eigenvalues of the relevant dynamical matrix coincide with the normal-mode frequencies, $\omega_k=\sqrt{\lambda_k}\propto{k}$: since these are equally spaced, the dynamics does not show dispersion. Moreover, if the chain's dynamical matrix is bisymmetric, i.e., the chain is persymmetric, it is able to transfer pulses between its ends with 100\,\% efficiency, making it a `perfect' transmission channel. Such a result was reported in Ref.~\cite{HerrmannS1981} for $n=3$, 4, 5 (the proof in the last case being too lengthy to be reported): here analytic expressions for the perfect-chain parameters are given for arbitrary $n$.

In Section~\ref{s.result} the formalism is introduced, the main theorem is stated, and its proof is given. Section~\ref{s.mass-spring} is devoted to the application of the theorem to the mass-spring chain, yielding the proper sizing of masses and elastic constants that make the chain dynamics periodic and perfectly transmitting. In Section~\ref{s.conclusions} the main outcomes of this paper are summarized and some perspective applications are mentioned. The analytical approach to the de\,Boor-Golub algorithm is reported in \ref{a.analytic}.

\section{Main result}
\label{s.result}

A Jacobi matrix of order $n$ is a symmetric tridiagonal matrix
\begin{equation}
\bm{A} = 
\begin{bmatrix}
	 a_1   & -b_1 &  0    &\cdots    &  0      \\
	-b_1   &  a_2 & -b_2  &          & \vdots  \\
	 0     & -b_2 &  a_3  &\ddots    &  0      \\
	\vdots &      &\ddots &\ddots    & -b_{n-1}\\
	 0     &\cdots&  0    & -b_{n-1} &  a_n    \\
\end{bmatrix}
\label{e.A}
\end{equation}
with nonzero off-diagonal entries $\{b_i\}$. One can assume $b_i>0$, as the sign of any of these entries can be changed by a similarity transformation: indeed, the diagonal matrix $\bm{S}_m=\{\delta_{ij}\sign\,(m{-}j)\}$, where $\sign(j)=\pm1$ for $j\ge0$ or $j<0$, respectively, produces a matrix $\bm{S}_m^{-1}\bm{A}\bm{S}_m$ identical to $\bm{A}$, but for the sign of $b_m$.

The matrix $\bm{A}$ is further assumed to be persymmetric, i.e., symmetric with respect to its antidiagonal,
\begin{equation}
 a_i = a_{n{+}1-i}
~,~~~
 b_i = b_{n-i} ~.
\label{e.abmirror}
\end{equation}

The main result of this paper is the determination of the $n$ independent matrix elements of $\bm{A}$ such that its given eigenvalues $\{\lambda_k\}_{k=0}^{n-1}$ are proportional to the sequence of the squares of $n$ successive integers, starting from zero.
With the above assumptions, this inverse problem is known to have one unique solution~\cite{Hochstadt1967,Hald1976}. Analogous results are known for eigenvalues in a linear sequence~\cite{ChristandlDEL2004,GladwellJW2004}, $\lambda_k\propto{k}$, a case relevant to the study of quantum-state transfer~\cite{KarbachS2005}, as well as in other cases~\cite{IgelnikS2011,OsteJ2017,Chu2019}.

\medskip

\noindent {\bf Theorem 1}.
Let $\bm{A}$ be the $n{\times}n$ bisymmetric matrix~\eqref{e.A} with entries
\begin{equation}
\begin{aligned}
 a_i&=n{-}1+4(i{-}1)(n{-}i) ~, &&i=1,\,...,\,n~,
\\
 b_i&=\sqrt{i\,(2i{-}1)\,(n{-}i)\,(2n{-}2i{-}1)} ~,&& i=1,\,...,\,n{-}1~.
\label{e.ex.aibi}
\end{aligned}
\end{equation}
Then $\bm{A}$ has the eigenvalues
\begin{equation}
 \lambda_k=2\,k^2 ~,~~~~~~~ k=0,\,...,\,n-1 ~.
\label{e.lambdak}
\end{equation}

\noindent {\em Proof}.~ This result can be proven by induction on $n$, so for clarity an explicit argument is used here for the matrix dimension. For $n\,{=}\,1$ the statement is trivially true, as $\bm{A}(1)= \big[\,0\,\big]$ has the eigenvalue $\lambda_0=0$. Assuming the statement to hold true for dimension $n$, one has to show that $\bm{A}(n{+}1)$ has the $n$ eigenvalues of $\bm{A}(n)$ plus the eigenvalue $\lambda_{n}=2n^2$. From Eq.~\eqref{e.ex.aibi} one has the entries of $\bm{A}(n{+}1)$,
\begin{equation}
\begin{aligned}
 a_i&=n+4(i{-}1)(n{+}1{-}i) ~, &&i=1,\,...,\,n{+}1~,
\\
 b_i&=\sqrt{i\,(2i{-}1)\,(n{+}1{-}i)\,(2n{-}2i{+}1)} ~,&&i=1,\,...,\,n~.
\end{aligned}
\label{e.ex.aibi1}
\end{equation}
Consider the tridiagonal matrix $\bm{C}(n{+}1)\coloneqq 2n^2\bm{I}-\bm{A}(n{+}1)$,
\begin{equation}
\bm{C} = 
\begin{bmatrix}
	 c_1   &  b_1 &  0    &\cdots  &  0     \\
	 b_1   &  c_2 &  b_2  &        & \vdots \\
	 0     &  b_2 &  c_3  &\ddots  &  0     \\
	\vdots &      &\ddots &~\ddots &  b_n   \\
	 0     &\cdots&  0    &  b_n   & c_{n+1} \\
\end{bmatrix}~,
\label{e.CN1}
\end{equation}
whose diagonal elements, from Eq.~\eqref{e.ex.aibi1}, are
\begin{equation}
 c_i=2n^2-a_i=n(n{-}1)+(n{+}2{-}2i)^2 ~,~~~~~ i=1,\,...,\,n{+}1~.
\label{e.ex.ci}
\end{equation}
One can factorize
\begin{equation}
 \bm{C} = \bm{H}\bm{H}^{\rm T} ~,
\end{equation}
where the matrix $\bm{H}$ is lower bidiagonal,
\begin{equation}
\bm{H} = 
\begin{bmatrix}
	 h_1   &  0   &  0    &\cdots   &  0      \\
	 r_1   &  h_2 &  0    &         & \vdots  \\
	 0     &  r_2 &  h_3  &\ddots   & \vdots  \\
	\vdots &      &\ddots &~\ddots  &  0      \\
	 0     &\cdots&  0    &  r_N    &  h_{n+1} \\
\end{bmatrix}~,
\label{e.HN1}
\end{equation}
its elements being given by
\begin{equation}
\begin{aligned}
 h_i&=\sqrt{(n{+}1{-}i)(2n{-}2i{+}1)} ~, &&i=1,\,...,\,n{+}1~,
\\
 r_i&=\sqrt{i\,(2i{-}1)} ~,&&i=1,\,...,\,n~.
\label{e.ex.hiri}
\end{aligned}
\end{equation}
The matrix $\bm{A}(n{+}1)=2n^2\bm{I}-\bm{H}\bm{H}^{\rm T}$ has the same spectrum of the matrix
\begin{equation}
 2n^2\bm{I}-\bm{H}^{\rm T}\bm{H} = 
\begin{bmatrix}
	    &          &   &  0      \\
	    & \bm{A}(n)&   &  \vdots \\
	    &          &   &  0      \\
	 0  &\cdots    & 0 &  2n^2   \\
\end{bmatrix}~.
\end{equation}
The tensor-product structure of this matrix entails that $\bm{A}(n{+}1)$ has the same $n$ eigenvalues of $\bm{A}(n)$, plus the eigenvalue $\lambda_{n}=2n^2$. \hfill $\square$

\medskip

The matrix entries~\eqref{e.ex.aibi} are seen to be $O(n)$ at the matrix borders and increasing up to $O(n^2)$ towards the matrix center, with an almost parabolic shape.

\section{Mass-spring chain model}
\label{s.mass-spring}

\begin{figure}
\begin{center}
\includegraphics[width=0.45\textwidth]{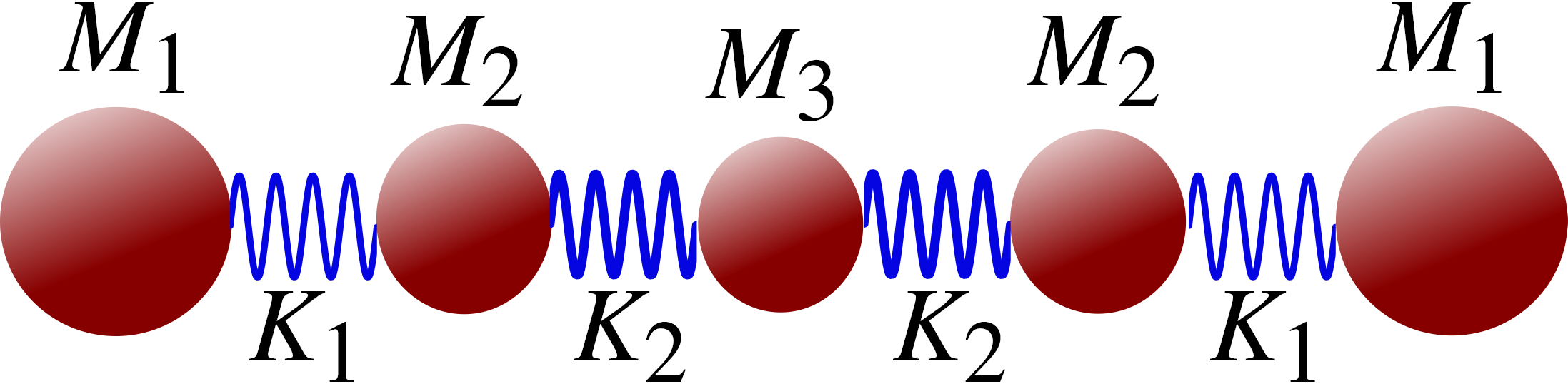}
\caption{A mirror-symmetric mass-spring chain with $n=5$ masses $(M_1,M_2,M_3,M_2,M_1)$ connected by 4 springs $(K_1,K_2,K_2,K_1)$.}
	\label{f.smc}
\end{center}
\end{figure}

One simple system where a Jacobi matrix comes into play is a mass-spring chain, see Fig.~\ref{f.smc}, consisting in a sequence of $n$ masses $\{M_i\}_{i=1}^n$ pairwise connected by $n{-}1$ ideal springs with elastic constants $\{K_i\}_{i=1}^{n-1}$; the neighboring masses $M_i$ and $M_{i+1}$ being coupled through the $K_i$ spring. This is quite a general model, as several different systems can be mapped onto it, such as electric LC circuits, ion chains, multilayered structures, and so on. Applications of the related dynamics can involve the study of energy transport in ion chains~\cite{RammPH2014} or synthesized nanostructures~\cite{CahillEtAl2003,NorrisLB2013}.

An intersting issue is the capability of the chain to transmit a pulse between the ends. It is well-known that a uniform chain, i.e., made of identical masses and springs, is a bad pulse transmitter, the worse the larger $n$: an initial pulse would undergo dispersion giving rise to a seemingly chaotic dynamics. In Ref.~\cite{Vaia2018}, using a result for quasi-uniform tridiagonal matrices~\cite{BanchiV2013}, it is shown how the transmission efficiency can be increased up to 98.7\,\% for any $n$, by properly tuning two extremal masses and their spring at both ends; it is indeed crucial to preserve the persymmetry of the dynamical matrix, equivalent to the chain's mirror-symmetry.
In the following it is shown that using the matrix defined in Theorem~1 one can design a mirror-symmetric mass-spring chain with 100\,\% transmission efficiency.

The mass-spring chain is characterized by a diagonal `mass' matrix,
\begin{equation}
\bm{M} =
\begin{bmatrix}
	 M_1   &  0   & \cdots &  0      \\
	 0     &  M_2 &        & \vdots  \\
	\vdots &      & \ddots    &  0      \\
	 0     &\cdots&      0 &    M_n
\end{bmatrix} ~,
\end{equation}
and a tridiagonal `spring' matrix,
\begin{equation}
 \bm{K} =
 \begin{bmatrix}
	 K_1 & -K_1       &  0            &\cdots \\[2mm]
	-K_1 & K_1{+}K_2 & -K_2       &\cdots \\[2mm]
	 0      & -K_2       & K_2{+}K_3 &       \\[2mm]
	\vdots  & \vdots        &               &\ddots
\end{bmatrix} ~;
\end{equation}
the eigenvalues of its dynamical matrix $\bm{M}^{-1/2}\bm{K}\,\bm{M}^{-1/2}$ are just the squares of the normal-mode frequencies of the chain,  $\big\{\omega_k\big\}_{k=0}^{n-1}$. Therefore, Theorem 1 has an immediate physical implication: by designing the chain such that
\begin{equation}
 \bm{M}^{-1/2}\bm{K}\,\bm{M}^{-1/2}=\frac{\omega^2}2\,\bm{A} ~,
\label{e.MKMA}
\end{equation}
then it turns out to have commensurate frequencies
\begin{equation}
 \omega_k = \omega\,k ~,~~~~~~~ k=0,\,...,\,n-1 ~,
\label{e.omegaklin}
\end{equation}
namely, integer multiples of a fixed frequency spacing $\omega$. Therefore, a perfectly periodic behavior with period $2\pi/\omega$ is determined, whatever the initial configuration. Moreover, as the chain is persymmetric ($M_i=M_{n{+}1-i}$, and $K_i=K_{n-i}$), the dynamical configuration after half a period $t^*=\pi/\omega$ becomes the mirror image of the initial one. This follows from the fact that the eigenvectors $v^{(k)}_i$ of $\bm{A}$ show alternate persymmetry~\cite{CantoniB1976},
\begin{equation}
 v^{(k)}_{N+1-i}=(-)^k\,v^{(k)}_i~;
\end{equation}
hence, denoting by $q_i(t)$ the displacement of the $i$-th mass at time $t$, the dynamics gives indeed~\footnote{The physical displacements are $\big\{q_i(t)/\sqrt{M_i}\big\}$, while $\{q_i(t)\}$ are the so-called `mass-weighted' displacements, see, e.g., Ref.~\cite{Vaia2018}.}
\begin{equation}
 q_i(t^*)=\sum_{k=0}^{n-1}v^{(k)}_i \cos(\omega k t^*) \sum_{j=1}^n v^{(k)}_jq_j(0)
         =q_{N+1-i}(0)
\end{equation}
as $\cos(\omega k t^*)=\cos(k\pi)=(-)^k$. For example, an elongation of the first mass is perfectly reproduced in the last mass after the time $t^*$, so the chain works as a perfect channel for, say, energy transmission.

The matrix equality~\eqref{e.MKMA} can be made explicit as a relation between the entries of $\bm{A}$ given in Eq.~\eqref{e.ex.aibi} and the chain parameters (assuming $K_0=K_n={0}$),
\begin{align}
\begin{aligned}
 \frac{\omega^2}2\,a_i &= \frac{K_i+K_{i-1}}{M_i}~,&~~~ &i=1,..., n~,
\\
 \frac{\omega^2}2\,b_i &= \frac{K_i}{\sqrt{M_iM_{i{+}1}}}~,& &i=1,\,...,\,n{-}1~.
\label{e.abKM}
\end{aligned}
\end{align}

\medskip

\noindent {\bf Theorem 2}.
 The elements of the matrices $\bm{M}$ and $\bm{K}$ can be reconstructed from the relation~\eqref{e.MKMA}, with the entries of $\bm{A}$ given by Eqs.~\eqref{e.ex.aibi}, to yield
\begin{equation}
\begin{aligned}
  M_{i+1} &= M_i\, \frac{2i{-}1}{i}\,\frac{n{-}i}{2n{-}2i{-}1} ~,
\\
 K_i &= \frac{M_i\omega^2}2\,(2i{-}1)(n{-}i) ~.
\end{aligned}
\label{e.MiKi1}
\end{equation}
The solution is unique once the value of $M_1$ is chosen.

\noindent {\em Proof}.
The proof that the solution exists and is unique is provided in Ref.~\cite{NylenU1997}. Setting for simplicity $\omega^2=2$, from Eqs.~\eqref{e.abKM} one has $K_i=\sqrt{M_iM_{i+1}}\,b_i$, and hence
\begin{equation}
 M_ia_i=K_i+K_{i-1}=\sqrt{M_iM_{i+1}}\,b_i+\sqrt{M_iM_{i-1}}\,b_{i-1}~.
\end{equation}
Defining the auxiliary variable
\begin{equation}
 x_i\coloneqq{b_i}\sqrt{\frac{M_{i+1}}{M_i}}=\frac{K_i}{M_i}
\label{e.xi}
\end{equation}
the above identity can be rearranged as a recursion for $x_i$,
\begin{equation}
 x_i=a_i-\frac{b_{i-1}^2}{x_{i-1}}~,
\end{equation}
starting with $x_1=a_1=n{-}1$; it is easy to check that the general solution is $x_i=(2i{-}1)(n{-}i)$, which directly gives the second of Eqs.~\eqref{e.MiKi1}, while the first one follows taking $(x_i/b_i)^2$, according to Eq.~\eqref{e.xi}.  \hfill $\square$

\medskip

Eqs.~\eqref{e.MiKi1} can be converted into closed expressions in terms of binomial coefficients,
\begin{equation}
\begin{aligned}
 M_i &= M_1\binom{n{-}1}{i{-}1}^2 ~\binom{2n{-}2}{2i{-}2}^{-1}~,
\\
 K_i &= M_1\omega^2(n{-}1)^2~ \binom{n{-}2}{i{-}1}^2
 ~\binom{2n{-}2}{2i{-}1}^{-1} ~.
\end{aligned}
\label{e.MiKi}
\end{equation}
The property of persymmetry, $M_i = M_{n{+}1-i}$ and $K_i = K_{n-i}$, is immediately verified. Clearly, a proportional scaling of masses and elastic constants does not affect the frequencies: this is reflected in the arbitrary choice of the first mass $M_1$. Since  the above expressions are rational numbers, by a proper choice of the parameters $M_1$ and $\omega$ the sequence of masses and elastic constants can be turned into positive `magic' coprime integers: these are reported in Table~\ref{t.magic} for $n\le10$.
From the expressions~\eqref{e.MiKi} it appears that for $i{+}1<n/2$ it is
\begin{equation}
\begin{aligned}
 \frac{M_{i+1}}{M_i}&=\frac{1-\frac1{2i}}{1-\frac1{2(n{-}i)}} <1 ~,
\\
 \frac{K_{i+1}}{K_i}&=\frac{1+\frac1{2i}}{1+\frac1{2(n{-}i{-}1)}} >1 ~,
\end{aligned}
\end{equation}
Therefore, the masses $M_i$ decrease and the elastic constants $K_i$ increase while moving towards the chain middle, where one finds the smallest mass(es) $M_{\lceil{n/2}\rceil}$ and the largest spring(s) $K_{\lceil{n/2}\rceil}$. By a Stirling approximation of the binomial coefficients for large $n$ one finds $M_{\lceil{n/2}\rceil}/M_1\simeq2/\sqrt{\pi{n}}$ and $K_{\lceil{n/2}\rceil}/K_1\simeq\sqrt{{n}/\pi}$, so that the `imbalance' of the elements in the mass and spring matrices, $\bm{M}$ and $\bm{K}$, is much smaller than in the matrix $\bm{A}$, for which the ratio between largest and smallest entries is $\simeq{n}$.

\begin{figure}
\begin{center}
\includegraphics[width=0.45\textwidth]{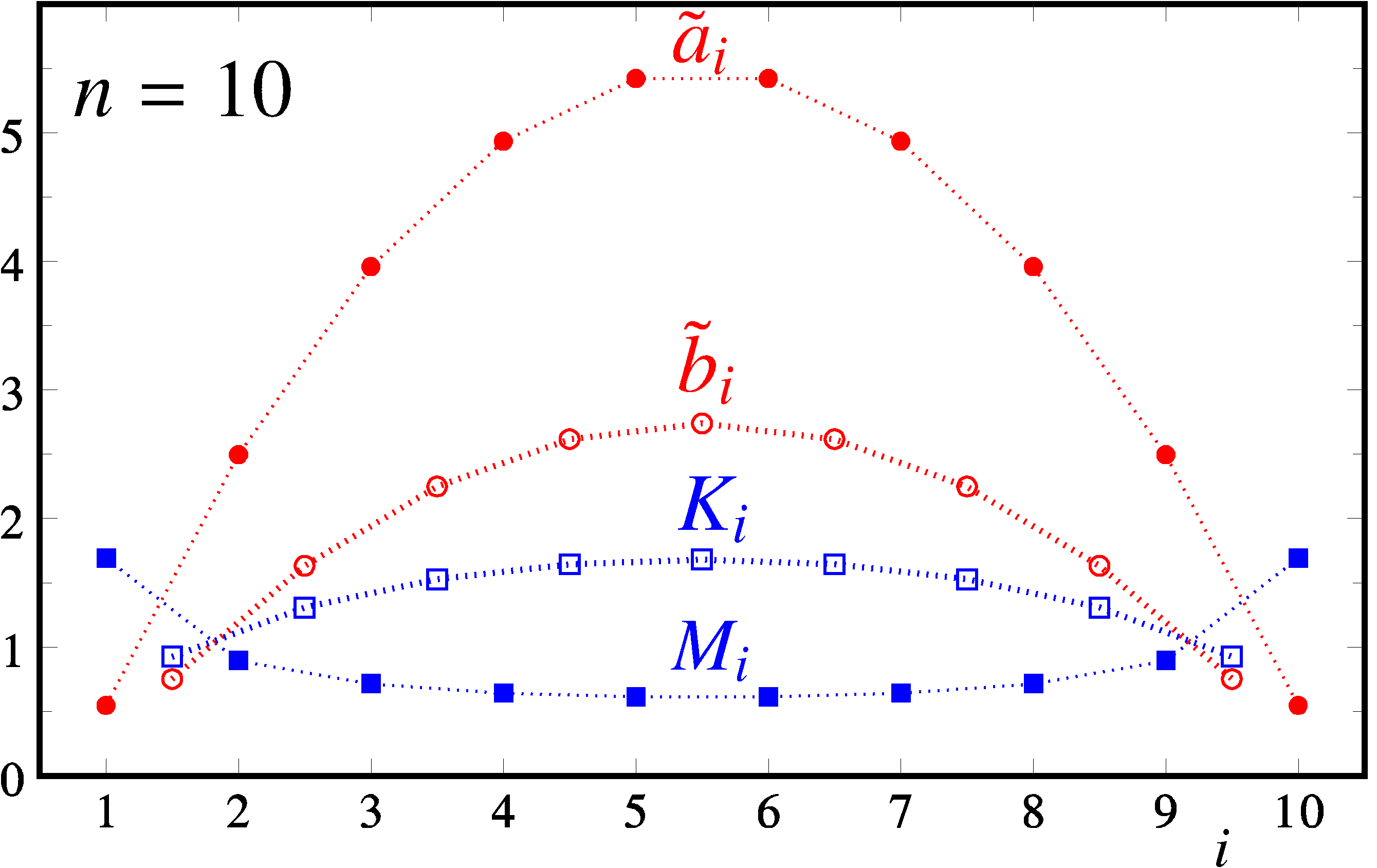}\\[1mm]
\includegraphics[width=0.45\textwidth]{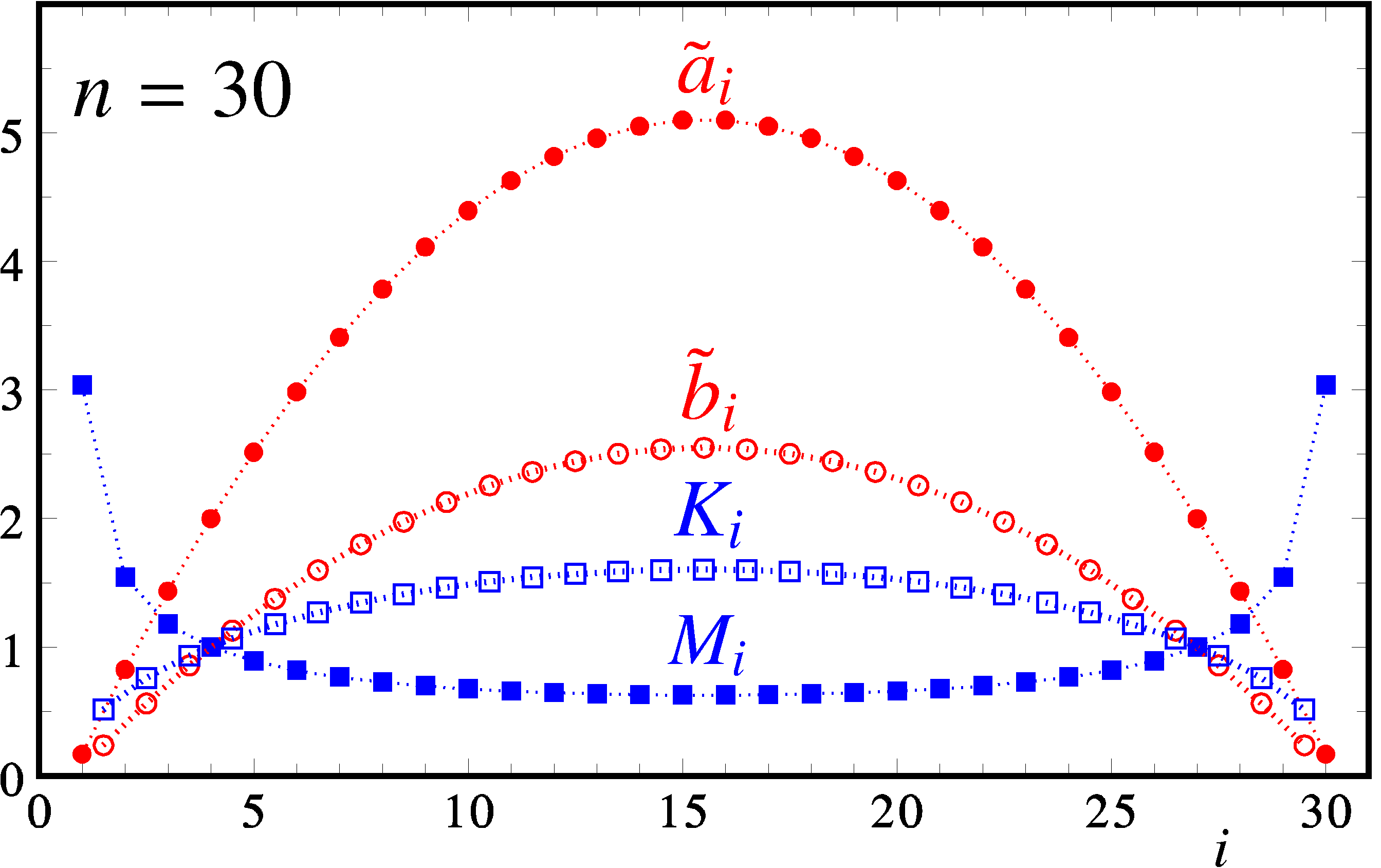}\\[1mm]
\includegraphics[width=0.45\textwidth]{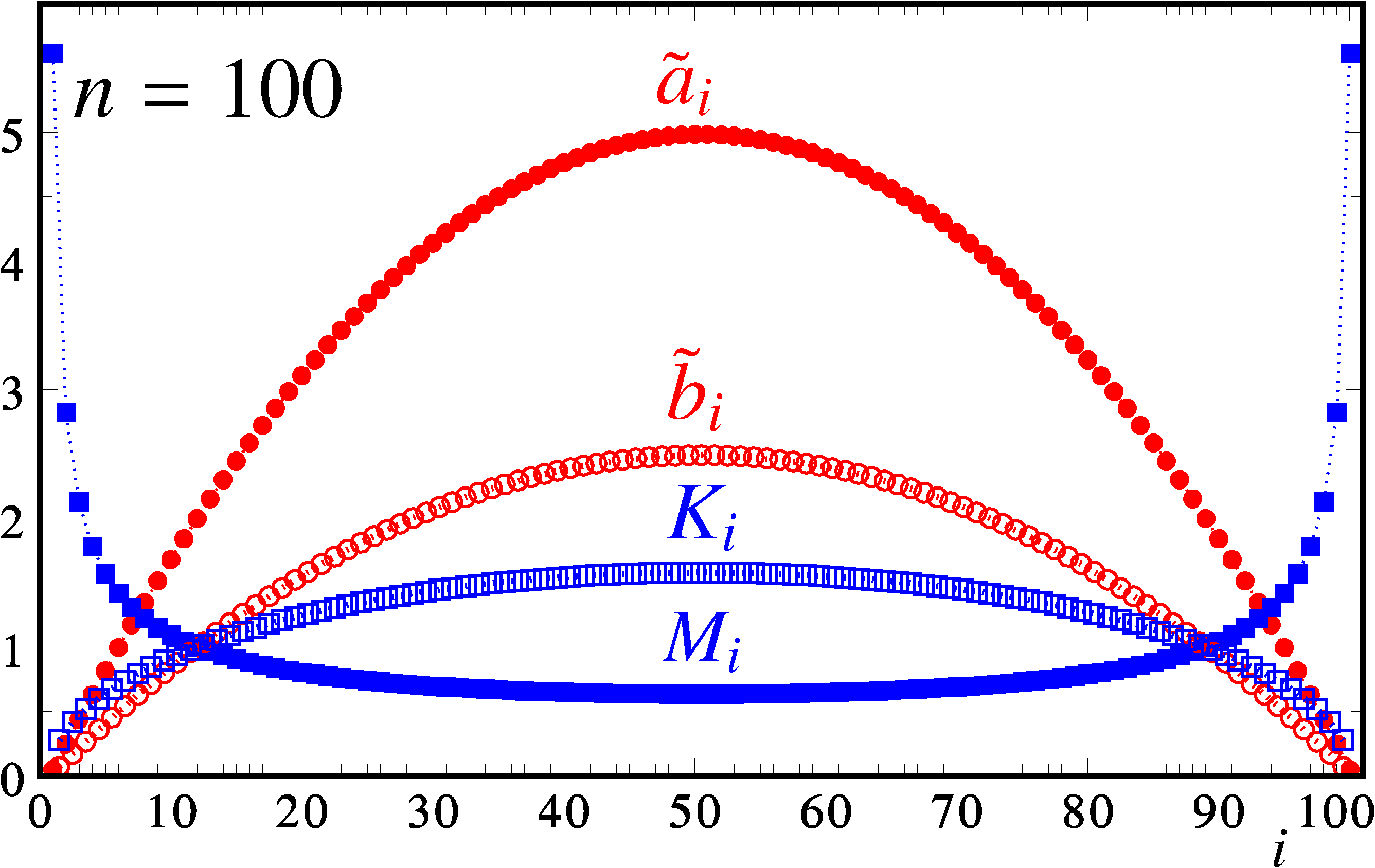}
\caption{Parameters of the matrix ${\bm{A}}$ (circles), see Eq.~\eqref{e.ab_ninfty}, and of the mass-spring chain (squares) for $n=10$, 30, and 100. The frequency spacing and the first mass are chosen as $\omega=\pi/(n{-}1)$ and $M_1\,{=}\,\sqrt{(n{-}1)/\pi}$. For a better visualization the $b_i$ and $K_i$ (open symbols), that regard the interaction between the sites $i$ and $i{+}1$, are put at the intermediate abscissa $i{+}\frac12$. Dotted lines are a guide for the eye.}
	\label{f.abKM}
\end{center}
\end{figure}

\begin{figure}
\begin{center}
\includegraphics[width=0.45\textwidth]{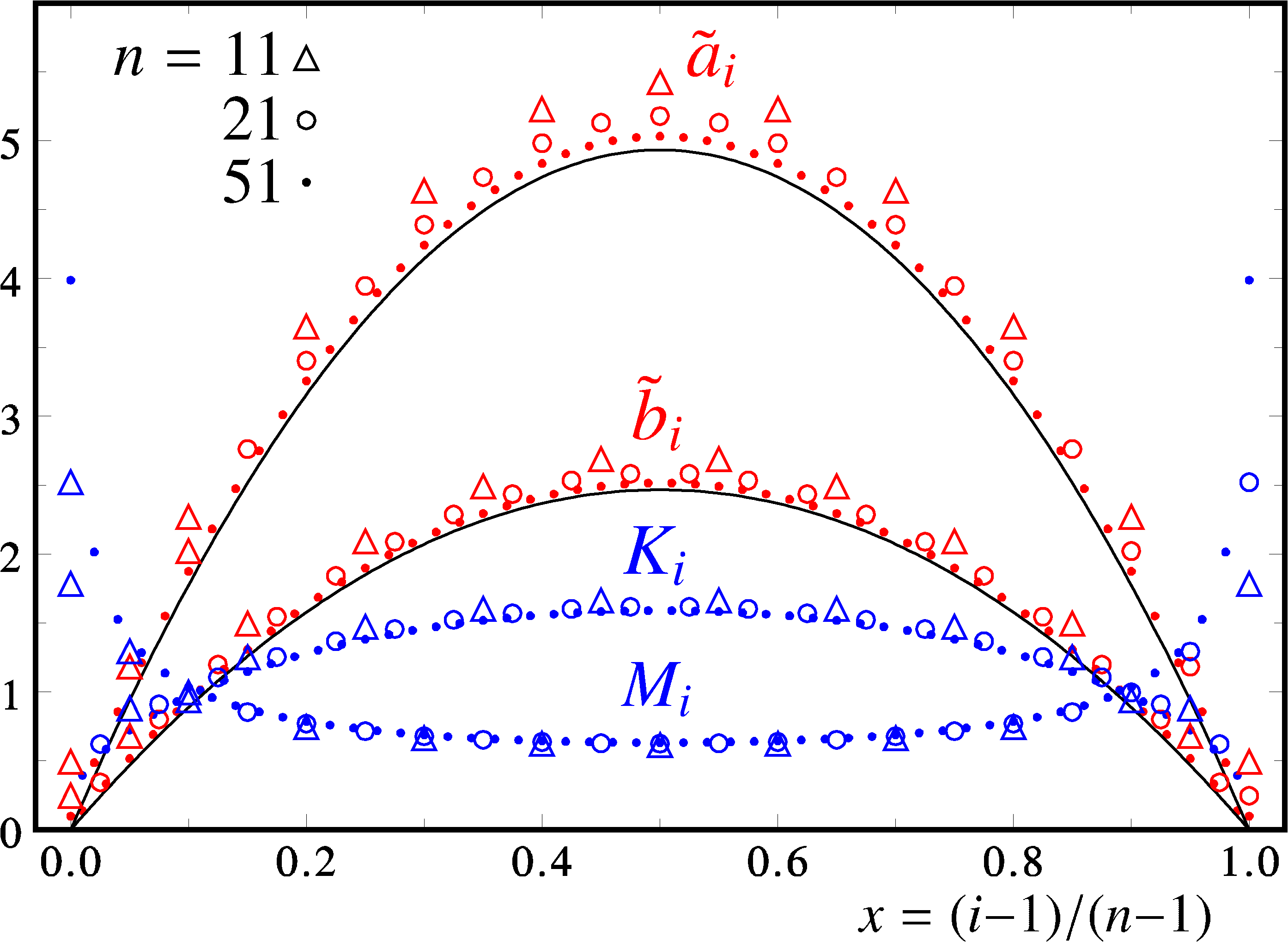}
\caption{Parameters of the matrix ${\bm{A}}$ (red symbols), see Eq.~\eqref{e.ab_ninfty}, and of the mass-spring chain (blue symbols) for $n=11$ (triangles), 21 (circles), and 51 (bullets). The frequency spacing and the first mass are chosen as $\omega=\pi/(n{-}1)$ and $M_1=\sqrt{(n{-}1)/\pi}$, entailing $M_{\lceil{n/2}\rceil}\to{2/\pi}$ and $K_{\lceil{n/2}\rceil}\to{\pi/2}$. The black curves report the parabolic shapes~\eqref{e.ab_ninfty} for the limit $n\to\infty$.}
	\label{f.collapse}
\end{center}
\end{figure}

A physically sound assumption is that the transmission time, $t^*=\pi/\omega$, scale linearly with the chain `length', $n{-}1$: without loss of generality one can choose the time unit such that $t^*=n{-}1$, i.e., $\omega=\pi/(n{-}1)$. With this choice, the behavior of the matrix entries and of the corresponding masses and elastic constants is shown, for different values of $n$, in Fig.~\ref{f.abKM}. From Eqs.~\eqref{e.ex.aibi} the large-$n$ shapes are described, in terms of the variable $x=(i{-}1)/(n{-}1)$, by parabolas,
\begin{equation}
\begin{aligned}
 \tilde a_i &\coloneqq \frac{\omega^2}2\,a_i ~~\longrightarrow~~  2\pi^2\,x\,(1-x) ~,
\\
 \tilde b_i &\coloneqq \frac{\omega^2}2\,b_i ~~\longrightarrow~~  \pi^2\,x\,(1-x)~,
\end{aligned}
\label{e.ab_ninfty}
\end{equation}
as confirmed by Fig.~\ref{f.collapse}; of course, the large-$n$ limit cannot account for the discrete values at the extrema.

\begin{figure}
\begin{center}
\includegraphics[width=0.45\textwidth]{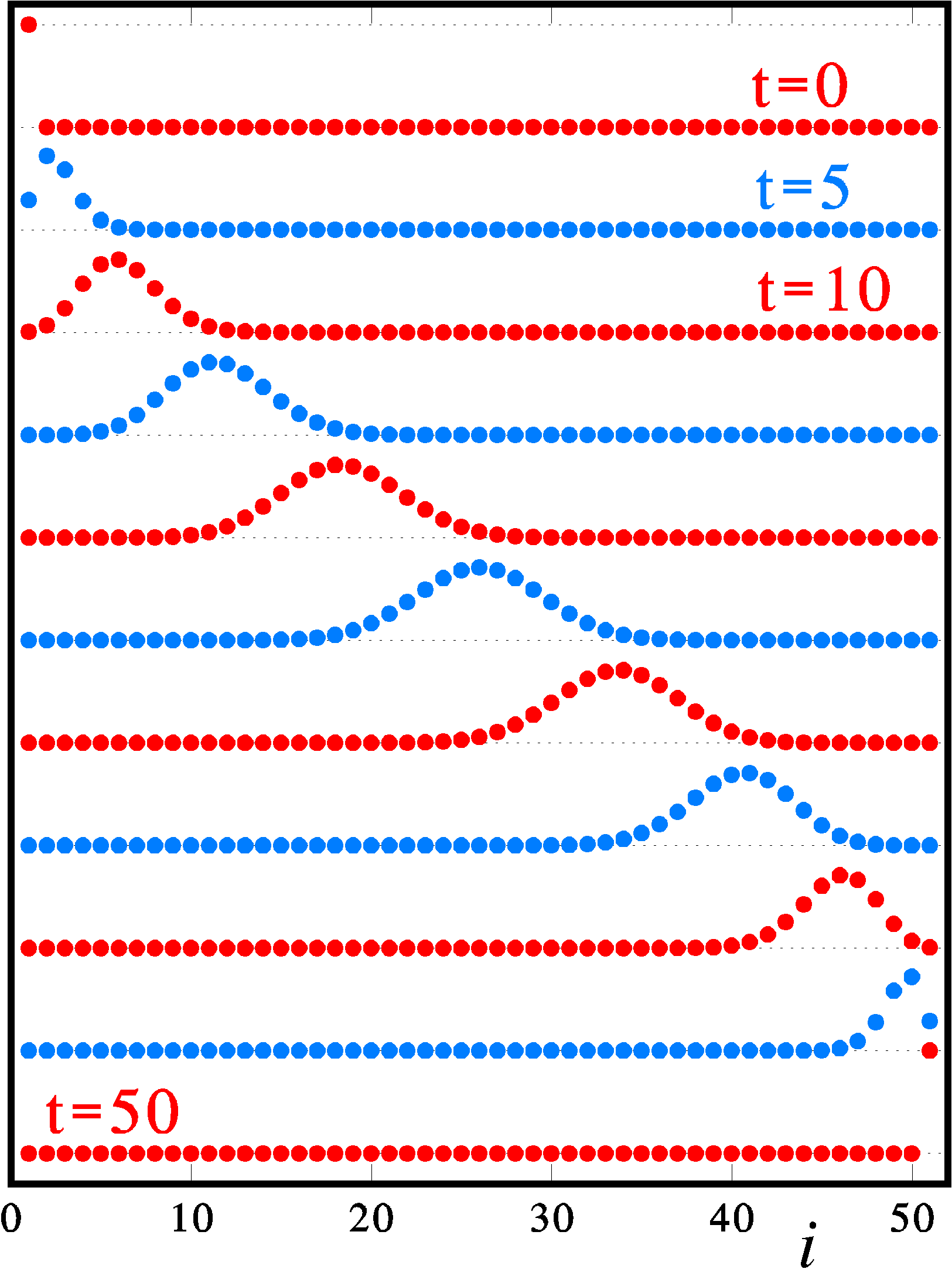}
\caption{Snapshots of the dynamics of a perfect chain with $n=51$ masses from time $t=0$ to $t^*=50$ at equal time intervals of 5 units. The abscissa labels the masses along the chain, while their displacements are represented as ordinates. It appears that the initial pulse spreads during its travel, become widest in the chain middle, then it shrinks and it perfectly recombines at $t^*$ to the mirror image of the initial configuration.}
	\label{f.d0051}
\end{center}
\end{figure}

\begin{figure}
\begin{center}
\includegraphics[width=0.45\textwidth]{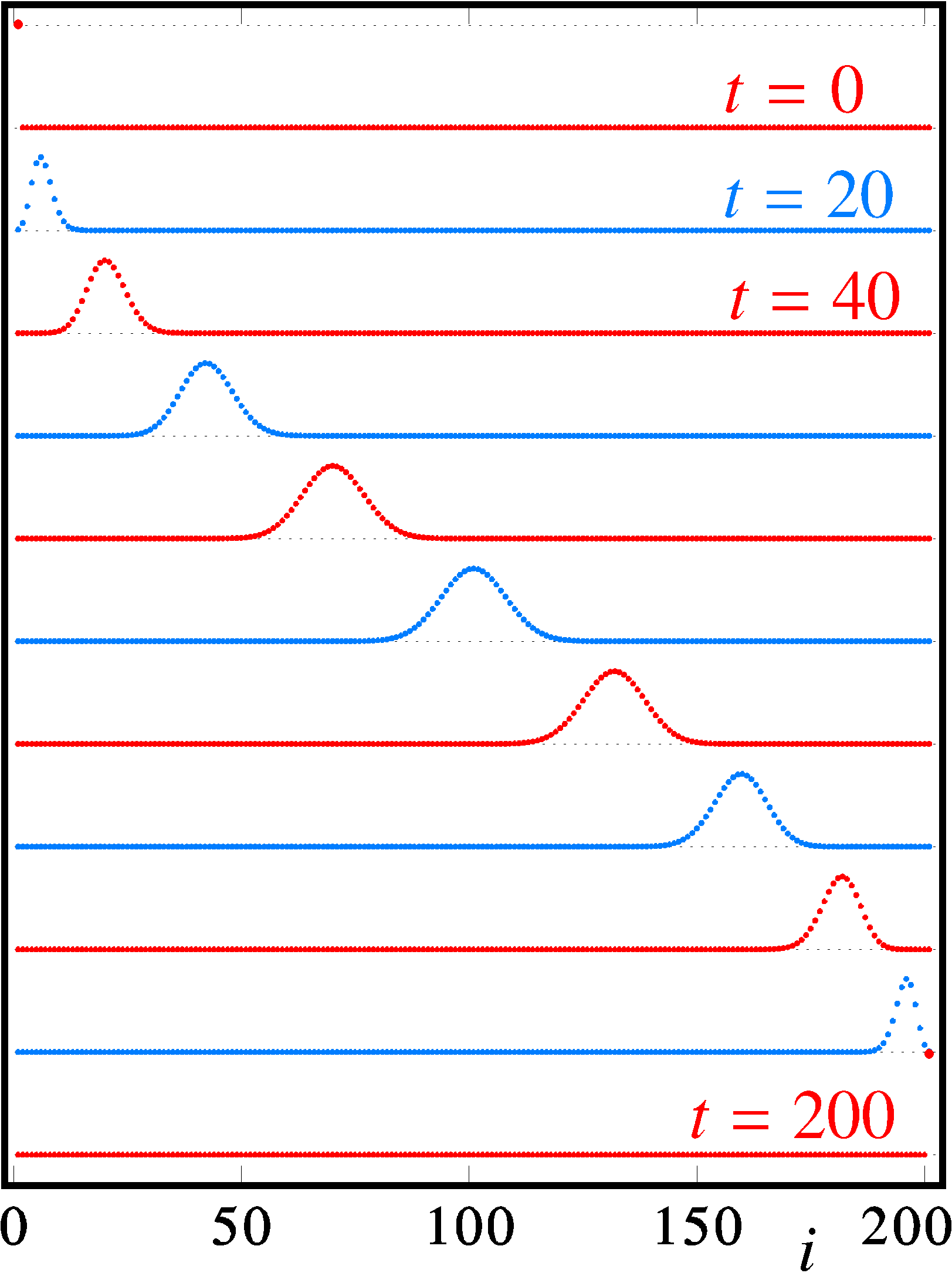}
\caption{Snapshots of the dynamics of a perfect chain with $n=201$ masses from time $t=0$ to $t^*=200$ at equal time intervals of 20 units.}
	\label{f.d0201}
\end{center}
\end{figure}

As remarked above, once the chain parameters are tuned as in Eqs.~\eqref{e.MiKi} the dynamics becomes perfectly periodic and dispersionless, as one can verify in Figs.~\ref{f.d0051} and~\ref{f.d0201}, where snapshots of the dynamics of chains with $51$ and $201$ masses are shown from $t=0$ until the half period $t=t^*$. Initially the chains are static, with the only first mass displaced from its equilibrium position, then a wider and wider traveling pulse is generated, involving simultaneous displacements of several masses; eventually, at $t^*=n{-}1$ it recomposes to a displacement of the only last mass, mirroring the initial shape. The subsequent evolution leads again to the initial configuration, and so on, periodically.

\section{Conclusions}
\label{s.conclusions}

In this paper a persymmetric Jacobi matrix of order $n$ has been analytically constructed, such that its eigenvalues are proportional to the squares of the first $n$ integers, $\lambda_k=\omega^2\,k^2,~(k=0,\,...,\,n{-}1)$.
Besides the usefulness for testing numerical algorithms devoted to solve inverse problems, such that introduced by de\,Boor and Golub~\cite{deBoorG1978}, the matrix has a practical interest as it enables one to design arbitrarily long perfectly periodic mass-spring chains. Indeed, the system frequencies are the square-roots of the eigenvalues, $\omega_k=\omega\,k,~(k=0,\,...,\,n{-}1)$, which are thus multiples of the frequency spacing $\omega$; hence, the dynamics returns to the initial configuration after a period $2\pi/\omega$. Moreover, thanks to mirror symmetry, at the time $t^*=\pi/\omega$ the configuration becomes the mirror image of the initial one, e.g., a deviation of the first mass is transferred to the last one with no dispersion. A basic application can be the realization of a toy behaving as a perfect Newton cradle, obtained by adding a pair of hanging masses hitting the ends of the chain (see Ref.~\cite{Vaia2018}): it would work also for a long chain, with the curious feature of the long duration of subsequent bounces. A more interesting application regards the transfer of energy pulses between the chain ends with perfect response, an effect that could lead, say, to efficient systems for heath or sound transmission. As the mass-spring model is an ubiquitous prototype for different, macroscopic and microscopic, systems studied in Physics, Engineering, Biology, and so on, the results presented in this paper are expected to be of potentially wide interest.

\bigskip\noindent
{\bf Acknowledgments}

R.V. thanks L.~Banchi, T.~J.~G.~Apollaro, A.~Cuccoli, and P.~Verrucchi for sharing precious comments.

\appendix
\section{Analytical approach to the de\,Boor-Golub algorithm}
\label{a.analytic}

The de\,Boor-Golub algorithm~\cite{deBoorG1978} is based on a sequence of $n{+}1$ monic polynomials $\{\chi_i(\lambda)\}$ of degree $i$, for $i=0,\,...,\,n$, corresponding to the characteristic polynomial $\chi_n(\lambda)$ of the matrix $\bm{A}$ and those of its left principal submatrices of order $i=0,\,...,\,n{-}1$. These polynomials are orthogonal under the internal product defined as
\begin{equation}
 \big\langle \chi, \tilde\chi \big\rangle 
 \coloneqq \sum_{k=0}^{n-1} w_k~ \chi(\lambda_k)~\tilde\chi(\lambda_k) ~,
\end{equation}
with weights given, if $\bm{A}$ is persymmetric, in terms of the eigenvalues as
\begin{equation}
 w_k = w\, \prod_{\substack{q=0\\q\ne{k}}}^{n-1}\frac1{\big|\lambda_k-\lambda_q\big|} ~,
\label{e.w}
\end{equation}
where $w$ is a nonzero constant. Then the polynomials can be sequentially constructed starting from $\chi_0(\lambda)=1$,
\begin{equation}
 \chi_{i+1}(\lambda)=(\lambda-a_{i+1})\,\chi_i(\lambda)-b_i^2\chi_{i-1}(\lambda) ~,
\label{e.chii}
\end{equation}
where it is assumed $b_0=0$ and the coefficients
\begin{equation}
 a_{i+1}=\frac{\big\langle \lambda\,\chi_i, \chi_i\big\rangle}
                 {\big\langle \,\chi_i, \chi_i\big\rangle}
~,~~~~~
 b_i^2=\frac{\big\langle \,\chi_i, \chi_i\big\rangle}
           {\big\langle \,\chi_{i-1}, \chi_{i-1}\big\rangle}
\label{e.ai_bi}
\end{equation}
correspond to the unique elements of the matrix~\eqref{e.A}. Note that $w$ can be arbitrarily chosen, as the above expressions do not depend on it.
The solution procedure, which turns out to be numerically very stable, starts by calculating and storing the weights~\eqref{e.w} and proceeds in a simple manner by iterating Eqs.~\eqref{e.chii} and~\eqref{e.ai_bi}.

Looking for a Jacobi matrix $\bm{A}$ having the eigenvalues~\eqref{e.lambdak}, the weights~\eqref{e.w} are given by (setting $m={n{-}1}$ for convenience)
\begin{equation}
\begin{aligned}
 \frac{w}{w_k} &=\prod_{\substack{q=0\\ q\ne k}}^m \big|k^2{-}q^2\big|
               =\prod_{q=0}^{k-1}(k{-}q)~\prod_{q=k+1}^m(q{-}k)
                ~\prod_{\substack{q=0\\ q\ne k}}^m(k{+}q)
\\
   &=k!~(m{-}k)!~\frac{(m{+}k)!}{(k{-}1)!}\,\frac1{2k}
    =\frac12~(m{-}k)!~(m{+}k)! ~.
\label{e.wkm1}
\end{aligned}
\end{equation}
Choosing $w=2m!/2^{2m{-}1}$ the weights take a compact expression in terms of a binomial coefficient,
\begin{equation}
 w_k=\frac1{2^{2m}}\binom{2m}{m{+}k} ~,
\label{e.w.n2}
\end{equation}
this result holding for $k\ne{0}$, because in~\eqref{e.wkm1} the exclusion of the $q{=}k$ term is accounted for by the denominator; the correct zeroth weight is directly given by
\begin{equation}
 \frac{w}{w_0}=\prod_{q=1}^m q^2 = (m!)^2 ~,
\end{equation}
so $w_0$ is half as that given by Eq.~\eqref{e.w.n2}: this can be accounted for by extending the summations to negative values of $k=-m,\,...,m$, with $w_{-k}=w_k$ according to Eq.~\eqref{e.w.n2}. Since
\begin{equation}
 \sum_{k=-m}^m w_k 
   = \frac1{2^{2m}}\sum_{k=-m}^m\binom{2m}{m{+}k} 
   = \frac1{2^{2m}}\sum_{k=0}^{2m}\binom{2m}{k} = 1 ~,
\end{equation}
the weights  $\big\{w_k\big\}_{k=-m}^m$ constitute a discrete probability distribution; denoting the averages with a double bracket, its characteristic function is
\begin{equation}
 \bigdave{e^{xk}}\coloneqq\sum_{k=-m}^m w_k\,e^{xk}
  = \cosh^{2m}\frac{x}2 ~,
\label{e.ch2mx}
\end{equation}
and by Taylor expansion one finds the moments
\begin{equation}
 \bigdave{k^{2\ell}} = 
  \bigg(\frac{d}{d x}\bigg)^{2\ell} \cosh^{2m}\frac{x}2 \,\bigg|_{x=0} ~.
\label{e.moments}
\end{equation}
Consider now the recursive algorithm of Eqs.~\eqref{e.chii} and~\eqref{e.ai_bi} and define the quantities
\begin{equation}
\begin{aligned}
 s_i&\coloneqq\big\langle \,\chi_i, \chi_i\big\rangle
     =\Bigdave{\,\chi_i^2(2k^2)} ~,
\\
 t_i&\coloneqq\big\langle \lambda\,\chi_i, \chi_i\big\rangle
     =\Bigdave{2k^2\,\chi_i^2(2k^2)}~,
\end{aligned}
\end{equation}
such that
\begin{equation}
 a_{i+1}=\frac{t_i}{s_i}
~,~~~~~~
 b_i^2=\frac{s_i}{s_{i-1}} ~.
\end{equation}
Being $\chi_0=1$ one immediately finds that $s_0=1$ and
\begin{equation}
 a_1=t_0=\bigdave{2k^2}=2\,\frac{d^2}{d x^2}\,\cosh^{2m}\frac{x}2\,\bigg|_{x=0}
    = m = n{-}1 ~,
\end{equation}
which agrees with~\eqref{e.ex.aibi}; then, as $\chi_1=\lambda-1$, it follows that $s_1=4\bigdave{k^4}-4\bigdave{k^2}^2$ and using Eq.~\eqref{e.moments} one can find
\begin{equation}
 b_1^2 = s_1 = m\,(2m-1) = (n-1)(2n-3) ~.
\end{equation}
With little effort one obtains $t_1=m\,(2m{-}1)\,(5m{-}4)$, yielding $a_2=m\,+4(m-1)=n-1+4(n-2)$; the analytical calculation of higher terms is increasingly cumbersome, but it soon leads to guessing Eqs.~\eqref{e.ex.aibi}, eventually proven in Section~\ref{s.result}.

\onecolumngrid

\renewcommand{\arraystretch}{1.1}
\setlength\tabcolsep{1.6mm}
\begin{table}
\caption{Perfectly transmitting chain: the `magic' sequences of integer masses $\{M_i\}_{i=1}^n$ and elastic constants $\{K_i\}_{i=1}^{n-1}$ for different chain lengths $n$. The second row reports the square of the corresponding frequency spacing $\omega$.}
\label{t.magic}
\medskip
\begin{center}
\begin{tabular}{|c|rr|rr|rr|rr|rr|rr|rr|rr|}
 \hline
  $n$ & \multicolumn{2}{c|}{3} &	\multicolumn{2}{c|}{4} 
	   & \multicolumn{2}{c|}{5} &	\multicolumn{2}{c|}{6} 
		& \multicolumn{2}{c|}{7} &	\multicolumn{2}{c|}{8} 
		 & \multicolumn{2}{c|}{9} &	\multicolumn{2}{c|}{10}
\\  \hline
 $\omega^2$ & \multicolumn{2}{c|}{$1/3$} & \multicolumn{2}{c|}{$2/3$} 
		& \multicolumn{2}{c|}{$1/10$} & \multicolumn{2}{c|}{$2/15$} 
		& \multicolumn{2}{c|}{$1/21$} &	\multicolumn{2}{c|}{$2/7$}
  		& \multicolumn{2}{c|}{$1/36$} &	\multicolumn{2}{c|}{$2/15$}
\\ \hline
   $i$ & \multicolumn{1}{c}{$M_i$} & \multicolumn{1}{c|}{$K_i$}
		& \multicolumn{1}{c}{$M_i$} & \multicolumn{1}{c|}{$K_i$}
		 & \multicolumn{1}{c}{$M_i$} & \multicolumn{1}{c|}{$K_i$}
		  & \multicolumn{1}{c}{$M_i$} & \multicolumn{1}{c|}{$K_i$}
		   & \multicolumn{1}{c}{$M_i$} & \multicolumn{1}{c|}{$K_i$}
 		    & \multicolumn{1}{c}{$M_i$} & \multicolumn{1}{c|}{$K_i$}
		     & \multicolumn{1}{c}{$M_i$} & \multicolumn{1}{c|}{$K_i$}
		      & \multicolumn{1}{c}{$M_i$} & \multicolumn{1}{c|}{$K_i$} \\
 \hline
 1& 3& 1& 5& 5& 35& 7& 63& 21& 231& 33& 429& 429& 6435&  715& 12155& 2431\\
 2& 2& 1& 3& 6& 20& 9& 35& 28& 126& 45& 231& 594& 3432& 1001&  6435& 3432\\
 3& 3&  & 3& 5& 18& 9& 30& 30& 105& 50& 189& 675& 2772& 1155&  5148& 4004\\
 4&  &  & 5&  & 20& 7& 30& 28& 100& 50& 175& 700& 2520& 1225&  4620& 4312\\
 5&  &  &  &  & 35&  & 35& 21& 105& 45& 175& 675& 2450& 1225&  4410& 4410\\
 6&  &  &  &  &   &  & 63&   & 126& 33& 189& 594& 2520& 1155&  4410& 4312\\
 7&  &  &  &  &   &  &   &   & 231&   & 231& 429& 2772& 1001&  4620& 4004\\
 8&  &  &  &  &   &  &   &   &    &   & 429&    & 3432&  715&  5148& 3432\\
 9&  &  &  &  &   &  &   &   &    &   &    &    & 6435&     &  6435& 2431\\
10&  &  &  &  &   &  &   &   &    &   &    &    &     &     & 12155&     \\
 \hline
	\end{tabular}
\end{center}
\end{table}
\twocolumngrid

\end{document}